\documentclass[aps,prl,10pt,twocolumn,showpacs,superscriptaddress]{revtex4-1}

\usepackage{amsmath}
\usepackage{latexsym}
\usepackage{amssymb}
\usepackage{bm}
\usepackage{graphics,epstopdf}
\usepackage{color}
\usepackage{hyperref}

\usepackage{newlfont}
\usepackage{amsfonts}
\usepackage{amsthm}
\usepackage{graphicx}
\usepackage{epsfig}

\newcommand{\ket}[1]{|{#1}\rangle}
\newcommand{\bra}[1]{\langle{#1}|}
\newcommand{\braket}[2]{\langle {#1} | {#2} \rangle}

\usepackage{times}

\usepackage[up]{subfigure}

\newcommand{\be}{\begin{equation}}
\newcommand{\ee}{\end{equation}}
\newcommand{\bc}{\begin{center}}
\newcommand{\ec}{\end{center}}
\newcommand{\bea}{\begin{eqnarray}}
\newcommand{\eea}{\end{eqnarray}}
\newcommand{\ba}{\begin{array}}
\newcommand{\ea}{\end{array}}

\begin{document}
\title{Potent Value and Potent Operator with Pre- and Postselected Quantum Systems}

\author{Arun Kumar Pati}
\email{akpati@hri.res.in}
\affiliation{Quantum Information and Computation Group,\\
Harish-Chandra Research Institute, HBNI, Chhatnag Road, Jhunsi,
Allahabad 211 019, India}





\date{\today}

\begin{abstract}
We introduce a novel concept which we call as potent value of system observable for  pre- and post-selected quantum states.
This describes, in general, how a quantum system affects the state of the apparatus during the time between 
two strong measurements corresponding to pre- and post-selections. The potent value can be realized for any 
interaction strength and for arbitrary coupling between the system and the apparatus observables. Most importantly,
potent values generalize and unify the notion of the weak values and modular values of observables in quantum theory.
Furthermore, we define a potent operator which describes the action of one system on the another and show that 
superposition of time-evolutions and time-translation machines are potent operators.
These concepts may find useful applications in quantum information processing and can lead to technological benefits.
\end{abstract}

\maketitle

{\it  Introduction.--}
The concept of weak measurement and weak value was first introduced by Aharonov-Albert-Vaidman
\cite{aha,av} while investigating the properties of a quantum system in a pre- and post-selected ensembles.
 If the system is weakly coupled to an apparatus, then upon post-selection of the system, the apparatus wave function is shifted by a weak value.
The weak value can have strange properties.
In particular, the weak value can be a complex number,
take values outside the spectrum of the observable being measured, and can be arbitrarily large.
 In particular, if one measures an observable $A$ of the system weakly,
with a pre-selected state $\ket{\psi}$ at time $t_i$ and post-selected state
$\ket{\phi}$ at time $t_f$, the value of the observable measured at time $t_i\leq t \leq t_f$ is
given by the weak value of observable, which is $\braket{\phi}{A|\psi}/\braket{\phi}{\psi}$.
Operationally the weak value describes how system observable couples to apparatus observable effectively between two strong 
measurements.
The concept of weak measurement has been generalized in various directions in recent years
\cite{sw1,ADL,sw2,nori,shik}
and have found numerous applications \cite{sandu,cho,wise1,mir,js,ams1,ams2,pdav,hof,eli,jd}. On foundational front, weak values 
have provided new methods to measure the wave function directly \cite{jsl1}, measure non-Hermitian operators \cite{pati}, and 
 derive Heisenberg uncertainty relation from classical uncertainty relation for complex random variables \cite{wu}, to name a few.

In addition to weak value, the concept of modular value of an observable for pre- and post-selected quantum system has been defined \cite{lev}.
When the coupling is not weak, the action of the pre- and post-selected system on the qubit is completely described by modular variable.
In the weak coupling limit, modular value becomes the weak value. It is also possible that without the weak coupling limit, the 
modular value can give rise to weak value. However, this method uses a qubit system as a pointer that couples to the original quantum system.

 In this paper we introduce the concept of potent values for system observable for a  pre- and post-selected quantum states.
This describes, in general, how a quantum system affects the state of the apparatus during the time between 
two strong measurements corresponding to pre- and post-selections. The potent values can be realized for any 
interaction strength and for arbitrary coupling between the system and the apparatus observables. Most importantly,
potent values generalize the notion of the weak values and modular values of observables in quantum theory.
For example, if the coupling is weak then the potent value becomes the weak value and for arbitrary coupling to qubit system
this yields the modular value.
The operational meaning of the potent value is that between the time interval of two strong measurements, it describes 
how one system affects the other system. The action of one system on the other is captured by the potent values of the observable of
the system of interest. We define yet another concept which is called as the potent operator. The potent operator completely describe 
the action of one system on the another. As an illustration, we show that the superposition of time-evolutions and the quantum 
time-translation machine are actually potent operators. This shows that the notion of potent operator may find useful
application similar to weak values in quantum technology.


{\it Potent value.--}
We start with a system which is preselected in the state $\ket{\psi_i} = \ket{\psi} \in {\cal H}_1$ and
allow it to interact with another quantum system (an apparatus) initially prepared in the state $\ket{\Phi} \in {\cal H}_2$.
The measurement can be realized using 
the interaction Hamiltonian
\begin{align}
 H_{int}= g \delta(t-t_0) A \otimes P,
\end{align}
where $g$ is the strength of the interaction that is sharply peaked at $t=t_0$, $A$ is an observable of the system and $P$ is that of the apparatus.
This is the von Neumann model of measurement when the coupling strength is arbitrary. Note that when $g$ is small, then we can realize the weak
measurement. The interaction Hamiltonian allows the system and apparatus to evolve as
\begin{align}
 \ket{\psi} \otimes \ket{\Phi}  & \rightarrow e^{-\frac{i}{\hbar}g A \otimes P} \ket{\psi} \otimes \ket{\Phi} \nonumber\\
 & \approx (I - \frac{i}{\hbar}g A \otimes P ) \ket{\psi} \otimes \ket{\Phi}.
\end{align}
After the weak interaction, we post-select the system in the state $\ket{\phi}$ with the
post-selection
probability given by $p= |\braket{\phi}{\psi}|^2 (1 + 2 g Im \langle A\rangle_w \langle P \rangle)$ with
$\langle P \rangle = \braket{\psi}{P |\psi }$. This yields the desired weak value of $A$ as
given by
\begin{align}
\langle A\rangle_w = A_w(\phi|\psi) = \frac{\braket{\phi}{A |\psi }}{\braket{\phi}{\psi} }.
\end{align}

After the post-selection the final state of the apparatus (unnormalized) is given by
\begin{align}
\ket{\Phi_f} =  e^{-\frac{i}{\hbar}g \langle A\rangle_w  P } \ket{\Phi}.
\end{align}
Thus, in the weak coupling limit the apparatus state is affected by the weak value of the system observable. In particular, 
if $P$ is the momentum operator, then the apparatus wave function in the position representation will be shifted by the real
part of the weak value. The counter intuitive result is that such a weak measurement leads to the shift of the pointer observable which
can lie far outside the eigenvalue of the observable $A$ if the pre- and post-selected states are almost orthogonal, i.e., $|\braket{\phi}{\psi}| << 1$.

Now, we ask if $g$ is not small how does the system affects the apparatus between two strong measurements. Specifically, what is the nature of 
the `strange value' corresponding to observable $A$ that affects the apparatus state for arbitrary coupling strength? The answer to this question is that for arbitrary 
interaction strength it is not the weak value, rather the `potent value' that affects the apparatus state.
Note that the general evolution for the system and the apparatus can be represented as 
\begin{align}
 \ket{\psi} \otimes \ket{\Phi}  & \rightarrow \sum_k \bra{k} e^{-\frac{i}{\hbar}g A \otimes P} \ket{\Phi} \ket{\psi} \otimes \ket{k} \nonumber\\
 &  = \sum_k A_k \ket{\psi} \otimes \ket{k}.
\end{align}
where $A_k =  \bra{k} e^{-\frac{i}{\hbar}g A \otimes P} \ket{\Phi}$ and $\ket{k}$ is an orthonormal basis for the apparatus.
The state of the apparatus, after the general interaction and upon post-selection of the system in the state $\ket{\phi}$, is given by 
\begin{align}
  \ket{\Phi_f} = N \sum_k \frac{\bra{\phi} A_k \ket{\psi}}{ \braket{\phi}{\psi}} \ket{k},
\end{align}
where $N$ is a normalization factor. The set of complex numbers defined below are potent values of the system observable, given by
\begin{align}
\langle A^{(k)} \rangle_p  =  A^{(k)}_p(\phi|\psi) = \frac{\bra{\phi} A_k \ket{\psi}}{ \braket{\phi}{\psi}}.
\end{align}
Therefore, the final state of the apparatus (second system) is given by
\begin{align}
  \ket{\Phi_f} = N \sum_k  \langle A^{(k)} \rangle_p  \ket{k}.
\end{align}
The effect of pre- and post-selected system on the apparatus state is completely described by a set of potent values $\langle A^{(k)} \rangle_p$ of
system observable. Since the final apparatus state can also be expressed as $\ket{\Phi_f} = \sum_k C_k \ket{k} $, the potent values of the system actually describe the
state of the apparatus completely. Thus, by measuring the potent values (following the method in Ref. \cite{pati}), we can determine the quantum state of the 
apparatus. This shows, under certain conditions (like pre-and post-selection), the state of one system depends on the strange potent values of another system with which
it might have interacted or entangled. This means that the wave function of a quantum system may not be a property of its own and thus,  
raises a deep question about the nature of quantum states.

As we will show in the sequel, the interesting observation is that {\it the potent values generalize and unify the notion of weak values and modular values}. 
For weak interaction potent values are related to weak values 
and for arbitrary interaction and qubit meter (apparatus) potent values are related to modular values.

 In the weak coupling limit $g << 1$, we can check that all potent values result in the weak value effectively, thus leading to the final state 
 of the apparatus as given by (up to normalization)
 \begin{align}
  e^{-\frac{i}{\hbar}g A_w(\phi|\psi)  P } \ket{\Phi}.
\end{align}
Note that in the weak interaction limit the potent values are given by

\begin{align}
 A_p^{(k)}(\phi|\psi) \approx \braket{k}{\Phi} e^{ - \frac{i}{\hbar} g A_w(\phi|\psi) P_w(k|\Phi) }
\end{align}
and hence the final state of the apparatus is now given by
\begin{align}
& \sum_k A_p^{(k)}(\phi|\psi) \ket{k} \approx \sum_k  \braket{k}{\Phi} e^{ - \frac{i}{\hbar}g A_w(\phi|\psi) P_w(k|\Phi) } \ket{k}  \nonumber\\
& \approx  e^{-\frac{i}{\hbar}g A_w(\phi|\psi)  P } \ket{\Phi}
 \end{align}
 It may be noted that the potent values depend on the weak value of the system observable and also on set of weak values for the apparatus. However, the 
 intermediate weak values of the apparatus are unobserved. 
Thus, in the weak coupling limit the effect of potent value on the apparatus is exactly same as the effect of weak value, i.e., the apparatus wave function 
(in position space) is shifted by the real part of the weak value.

 Next we will show that for qubit meter, the potent values actually describe the modular values \cite{lev}. To show this consider a Hamiltonian $H = g A \otimes \Pi$ 
 that couples a system and a qubit with $\Pi= \ket{1}\bra{1}$. The initial state of the qubit $\ket{\Phi} = \alpha \ket{0} + \beta \ket{1}$, where 
 $\ket{0}, \ket{1}$ are the eigenbasis of the Pauli matrix $\sigma_z$. The unitary operator 
 that evolves system and the meter qubit is given by 
 \begin{align}
 U = e^{ - \frac{i}{\hbar} g A \otimes \Pi } = I + (e^{ - \frac{i}{\hbar} g A} - I ) \otimes \Pi
\end{align}
and the potent values are given by 
\begin{align}
 & A_p^{(0)}(\phi|\psi) = \frac{\bra{\phi} A_0 \ket{\psi}}{ \braket{\phi}{\psi}} = \alpha,  \nonumber\\
 & A_p^{(1)}(\phi|\psi) = \frac{\bra{\phi} A_1 \ket{\psi}}{ \braket{\phi}{\psi} } = \beta \frac{\bra{\phi} e^{ - \frac{i}{\hbar} g A}  \ket{\psi}}{ \braket{\phi}{\psi} } .
\end{align}
Therefore, using Eq.(8), the final state of the qubit is given by (up to normalization)
\begin{align}
\ket{\Phi_f} & = \langle A \rangle_p^{(0)}  \ket{0} + \langle A \rangle_p^{(1)}  \ket{1} \nonumber\\
& = \alpha \ket{0} +  \beta \frac{\bra{\phi} e^{ - \frac{i}{\hbar} g A}  \ket{\psi}}{ \braket{\phi}{\psi} }  \ket{1} \nonumber\\
& = \alpha \ket{0} + \beta  \langle A \rangle_M \ket{1},
\end{align}
where $ \langle A \rangle_M = \frac{\bra{\phi} e^{ - \frac{i}{\hbar} g A}  \ket{\psi}}{ \braket{\phi}{\psi} }  $ is called as the modular value of the 
observables $A$. It should be noted that the above equation is same as Eq(15) of Ref.\cite{lev} which has been used to define the modular value.
Similar to the weak value, the modular value is defined only for post-selected states which are not orthogonal to the pre-selected state. 
Thus, potent values unify the concept of weak values and modular values between two strong measurements. They completely describe the 
effect of pre- and post-selected quantum system on another system which may be qubit or any general quantum system.

 {\it Potent operator.--}
 For a pre- and postselected quantum system, one can define another concept which we call potent operator that acts on the apparatus Hilbert space and 
 describes the action of system on the apparatus. The system and apparatus evolves under unitary evolution as 
 $$\ket{\psi} \otimes \ket{\Phi} 
 \rightarrow U \ket{\psi} \otimes \ket{\Phi} = e^{-\frac{i}{\hbar}g A \otimes P} \ket{\psi} \otimes \ket{\Phi},$$
 where $g$ is an arbitrary coupling strength. Now, upon post-selection of system state $\ket{\phi}$, 
 the final (unnormalized) state of the apparatus is described by
 \begin{align}
\ket{\Phi_f}  = \frac{\bra{\phi} e^{ - \frac{i}{\hbar} g A \otimes P}  \ket{\psi}}{ \braket{\phi}{\psi} } \ket{\Phi} =  U_P(\phi|\psi) \ket{\Phi},
\end{align}
where  
 \begin{align}
 U_P(\phi|\psi) =  \frac{\bra{\phi} e^{ - \frac{i}{\hbar} g A \otimes P}  \ket{\psi}}{ \braket{\phi}{\psi} } 
 \end{align}
  is the potent operator that acts on the apparatus Hilbert space ${\cal H}_2$. Note that the potent operator is not a unitary operator. If the pre-selected state 
  of the system is one of the orthonormal basis of some observable (that does not commute with $A$), i.e., $\ket{\psi} = \ket{\psi_n}$ then, 
  the potent operator satisfies the identity
 \begin{align}
 \sum_n |\braket{\phi}{\psi_n}|^2  U_P(\phi|\psi_n)  U_P(\phi|\psi_n)^{\dagger} = I.
 \end{align}
  
   The potent operator has interesting connection to weak values and modular values. Suppose the joint unitary operator is a conditional unitary operator 
where system is a controlled one and apparatus is a target one. Then, we have $U = \sum_n \Pi_n \otimes U_n$ where $\Pi_n$'s are
projections on the system Hilbert space with $\Pi_n \Pi_m = \Pi_n \delta_{nm}$ and $U_n$'s are the set of unitaries that act on the apparatus Hilbert space. 
The potent operator now is given by
\begin{align}
 U_p(\phi|\psi) =  \sum_n \frac{\bra{\phi} \Pi_n \ket{\psi}}{ \braket{\phi}{\psi}} U_n.  
 \end{align}
In this case, the potent operator depends on the set of weak values corresponding to system observable as well as on the local unitary operators 
$U_n$. The final (unnormalized) state of the apparatus is then given by
\begin{align}
  \ket{\Phi_f} =  \sum_n  \langle \Pi_n \rangle_w U_n  \ket{\Phi}.
\end{align}

On the other hand if we have
a conditional unitary operator 
where apparatus is a controlled one and system is a target one, then we have $U = \sum_n V_n \otimes P_n$ where $V_n$'s are
unitary operators on the system Hilbert space and $P_n$'s are projectors on the apparatus Hilbert space with $P_n P_m= P_n \delta_{nm}$. 
In this case, the potent operator is given by
\begin{align}
 U_P(\phi|\psi) =  \sum_n \frac{\bra{\phi} V_n \ket{\psi}}{ \braket{\phi}{\psi}} P_n.  
 \end{align}
Since every unitary can be written as $V_n = e^{-i \lambda A_n}$ for some real parameter $\lambda$ and set of Hermitian operators $A_n$, we have 
\begin{align}
 U_P(\phi|\psi) =  \sum_n \frac{\bra{\phi} e^{-i \lambda A_n}  \ket{\psi}}{ \braket{\phi}{\psi}} P_n.  
 \end{align}
In this case, the potent operator depends on the set of modular values of system observables as well as the set of projectors 
that act on the apparatus Hilbert space. 
The final (unnormalized) state of the apparatus is then given by
\begin{align}
  \ket{\Phi_f} =  \sum_n  \langle A \rangle_M P_n  \ket{\Phi},
\end{align}
where $ \langle A_n \rangle_M = \frac{\bra{\phi} e^{-i \lambda A_n}  \ket{\psi}}{ \braket{\phi}{\psi}} $ is the set of modular values for the system
observable $A_n$.

Thinking in a more general setting, one can say
that between two strong measurements of system and apparatus, a conditional unitary dynamics shows that system acts on the 
apparatus via a set of weak values and the apparatus acts on the system via a set of modular values. Thus, modular operator provides
a new perspective on the role of weak values and modular values. Note that weak values appear here without weak interaction and 
modular values appear without any coupling parameter between system and apparatus observables. Moreover, the modular values appear 
here without the apparatus being a qubit system. Thus, potent operator may capture 
weak values and modular values beyond their region of validity.

 To see the potent operator for qubit meter, consider again the Hamiltonian  $H = g A \otimes \Pi$ with
 $\Pi= \ket{1}\bra{1}$ that couples a quantum system and a qubit. The initial state of the qubit $\ket{\Phi} = \alpha \ket{0} + \beta \ket{1}$.
 The potent operator that acts on the state of the meter qubit is given by 
 \begin{align}
 U_P(\phi|\psi) & = (I - \Pi) + \frac{\bra{\phi} e^{ - \frac{i}{\hbar} g A}  \ket{\psi}}{ \braket{\phi}{\psi} }  \Pi \nonumber\\
 & = \ket{0}\bra{0} + \langle A \rangle_M \ket{1}\bra{1}.
\end{align}
Thus, the potent operator that acts on a qubit meter depends on the modular value $\langle A \rangle_M$ of the system observable.
Therefore, the action of the potent operator on the apparatus state leads to the final state (unnormalized) 
\begin{align}
\ket{\Phi_f}  = \alpha \ket{0} + \beta  \langle A \rangle_M \ket{1}
\end{align}
which is same as in Eq.(14). This brings out another connection between the potent operator and the modular value.

{\it Superposition of time evolutions and potent operator.--}
In quantum theory, it is possible to have superpositions of several time-evolution operators $U_i$ which may effectively lead to a single time evolution operator \cite{aapv}
$U'$, i.e., one can have 
\begin{align}
\sum_i c_i U_i \ket{\Phi} \sim U' \ket{\Phi}
\end{align}
with $\sum_i c_i =1$. 
Though, this may not hold for all states, for certain systems it was shown that (24) can hold. If we consider a quantum system with Hamiltonian $H = H(a_i)$ and allows 
this to evolve for a time $T$, then the superposition of evolution with different $H(a_i)$'s can be expressed as 
\begin{align}
 \sum_i c_i U(T,a_i) \ket{\Phi} =  \sum_i c_i e^{-\frac{i}{\hbar} \int_{t_0}^{t_0+T} H(a_i) dt} \ket{\Phi}
\end{align}
It was shown by Aharonov-Anandan-Popescu-Vaidman (AAPV) \cite{aapv} that this superposition can lead to a single time evolution with a parameter $a'$ which 
can be far out of the range of the values given by $a_i$'s. The effective equation then is given by
\begin{align}
\sum_i c_i U(T,a_i) \ket{\Phi} \sim U(T,a') \ket{\Phi}.
\end{align}

We show that superposition of time evolution operator is actually a potent operator.
Let us consider a composite quantum system initially at $t=t_0$ in the state $\ket{\psi} \otimes \ket{\Phi}$ and the Hamiltonian $H= H(A,P)$ of the composite system 
is a function of a conserved quantum observable $A$ of one of the system with $A\ket{a_i} = a_i \ket{a_i}$. Let the initial state of the system be $\ket{\psi} = \sum_i c_i \ket{a_i}$. 
Then, under the unitary time-evolution for a period of time $T$, we have 
\begin{align}
 \ket{\psi} \otimes \ket{\Phi}  \rightarrow   \sum_i c_i e^{-\frac{i}{\hbar} \int_{t_0}^{t_0+T} H(a_i) dt} \ket{a_i} \ket{\Phi}
\end{align}
The potent operator for the apparatus state with pre- and postselected quantum system is given by
\begin{align}
 U_P(\phi|\psi) = \frac{\bra{\phi} e^{-\frac{i}{\hbar} \int_{t_0}^{t_0+T} H(a_i) dt} \ket{\psi}}{ \braket{\phi}{\psi}} 
\end{align}
If we choose the postselected state $\ket{\phi} = \frac{1}{\sqrt N} \sum_i \ket{a_i}$, then the potent operator is given by
\begin{align}
 U_P(\phi|\psi) =  \sum_i c_i e^{-\frac{i}{\hbar} \int_{t_0}^{t_0+T} H(a_i) dt} 
\end{align}
Using Eq.(15), we can see that the 
\begin{align}
\ket{\Phi} \rightarrow \ket{\Phi_f}=  U_P(\phi|\psi) \ket{\Phi} =  \sum_i c_i  e^{-\frac{i}{\hbar} \int_{t_0}^{t_0+T} H(a_i) dt} \ket{\Phi}
\end{align}
Thus, superpositions of several time-evolutions is actually a potent operator which results in a single time-evolution of other quantum system depending on the parameter 
which may be far out side the range of spectrum of the apparatus observable. 

  
 If one use the same Hamiltonian for different periods of time then using the Aharonov-Anandan-Popescu-Vaidman (AAPV) proposal \cite{aapv} one can construct a quantum time 
 translation machine. This utilizes the idea that the superposition of the time evolutions during different periods of time $T_i$ may result is the time evolution during 
 a period $T' = \sum_i c_i T_i$, which may be very different from the range of $\{T_i\}$. Curiously, this may result in a single time evolution towards the past or towards the future
 depending on whether the resultant $T'$ is negative or positive, respectively. This has no classical analog \cite{vaid}. 
 Again, one can show that the potent operator results in the time translation machine with suitable 
 pre- and postselection. This shows the power of potent operator.
 
{\it Conclusions.--}
  To summarize, in this paper we have introduced two new notions: the potent value for the system observable and the potent operator that can act on the apparatus state 
  for pre- and post-selected quantum system. The potent value generalizes the notion of 
  weak value and modular value for quantum observable and describes the action of one system on the other between two strong measurements of the former. 
  Thus, the concept of potent value unifies the weak and modular values. Also, we have argued that by measuring the potent value one can measure the wave function of
  a quantum system. The wave function of one system is described by the potent value of another system that has been pre- and post-selected. 
  This raises a deep question about the nature 
  of quantum states. Because, this shows that under certain condition wave function of a quantum system may not be a property of its own. Similarly, the concept of 
  potent operator describe the action of one system on the another and can have curious properties. The power of potent operator can be seen from the fact that this realizes the 
  notion of superposition of time-evolution operator and quantum time-translation machine. We hope that these new concepts may lead to novel effects which
  may have technological benefits 
  in quantum information and quantum theory in general.

\vskip 1cm
{\em Note:} These ideas were conceived during 2013 and paper was completed in 2014. I thank L. Vaidman and J. Dressel for useful discussions during 
the meeting on ``Concepts and Paradoxes in a Quantum Universe'' at Perimeter Institute, Waterloo during June 20-24, 2016. Finally, this paper is on arXiv 
in 2018.


\end{document}